\documentclass[aps,twocolumn]{revtex4}

\usepackage[latin1]{inputenc}
\usepackage{graphicx}

\begin{document}

\title{Wrinkling hierarchy in constrained thin sheets from suspended graphene to curtains}

\author{Hugues Vandeparre$^{1,}$\footnote{These authors contributed equally to this work}}
\author{Miguel Pi\~neirua$^{2,*}$}
\author{Fabian Brau$^1$}
\author{Benoit Roman$^2$}
\author{Jos\'e Bico$^2$}
\author{Cyprien Gay$^3$}
\author{Wenzhong Bao$^4$}
\author{Chun Ning Lau$^4$}
\author{Pedro M. Reis$^5$}
\author{Pascal Damman$^1$}
\email{pascal.damman@umons.ac.be}

\affiliation{$^1$Laboratoire Interfaces $\&$ Fluides Complexes, CIRMAP, Universit\'e de Mons, 20 Place du Parc, B-7000 Mons, Belgium}
\affiliation{$^2$PMMH, CNRS UMR 7636, ESPCI, ParisTech, Univ. Paris 6 \& Paris 7, 10 Rue Vauquelin, 75231 Paris Cedex 05, France}
\affiliation{$^3$Mati\`ere et Syst\`emes Complexes, Universit\'e Paris Diderot - Paris 7, CNRS, UMR 7057, B\^atiment Condorcet, F-75205 Paris cedex 13, France}
\affiliation{$^4$Department of Physics and Astronomy, University of California, Riverside, California 92521, USA}
\affiliation{$^5$Departments of Mechanical Engineering and Civil \& Environmental Engineering, Massachusetts Institute of Technology, Cambridge, Massachusetts 02139, USA}

\date{\today}

\begin{abstract}
We show that thin sheets under boundary confinement spontaneously generate a universal self-similar hierarchy of wrinkles. From simple geometry arguments and energy scalings, we develop a formalism based on \emph{wrinklons}, the transition zone in the merging of two wrinkles, as building-blocks of the global pattern. 
Contrary to the case of crumple paper where elastic energy is focused, this transition is described as smooth in agreement with a recent numerical work \cite{benny11}. 
This formalism is validated from hundreds of nm for graphene sheets to meters for ordinary curtains, which shows the universality of our description. 
We finally describe the effect of an external tension to the distribution of the wrinkles.
\end{abstract}

\maketitle

The drive towards miniaturization in technology is demanding for increasingly thinner components, raising new mechanical challenges~\cite{rogers}. Thin films are however unstable to boundary or substrate-induced compressive loads: moderate compression results in regular wrinkling~\cite{hutch,cerd03,poci08,bao09,brau10} while further confinement can lead to crumpling~\cite{Witten07,audo10}. Regions of stress focusing can be a hindrance, acting as nucleation points for mechanical failure. Conversely, these deformations can be exploited constructively for tunable thin structures. For example, singular points of deformation dramatically affect the electronic properties of graphene~\cite{pereira10}. 

Here, we show that thin sheets under boundary confinement spontaneously generate a universal self-similar hierarchy of wrinkles; from strained suspended graphene to ordinary hanging curtains. We develop a formalism based on \emph{wrinklons}, a localized transition zone in the merging of two wrinkles, as building-blocks to describe these wrinkled patterns. 

\begin{figure}
\includegraphics[width=0.76\columnwidth]{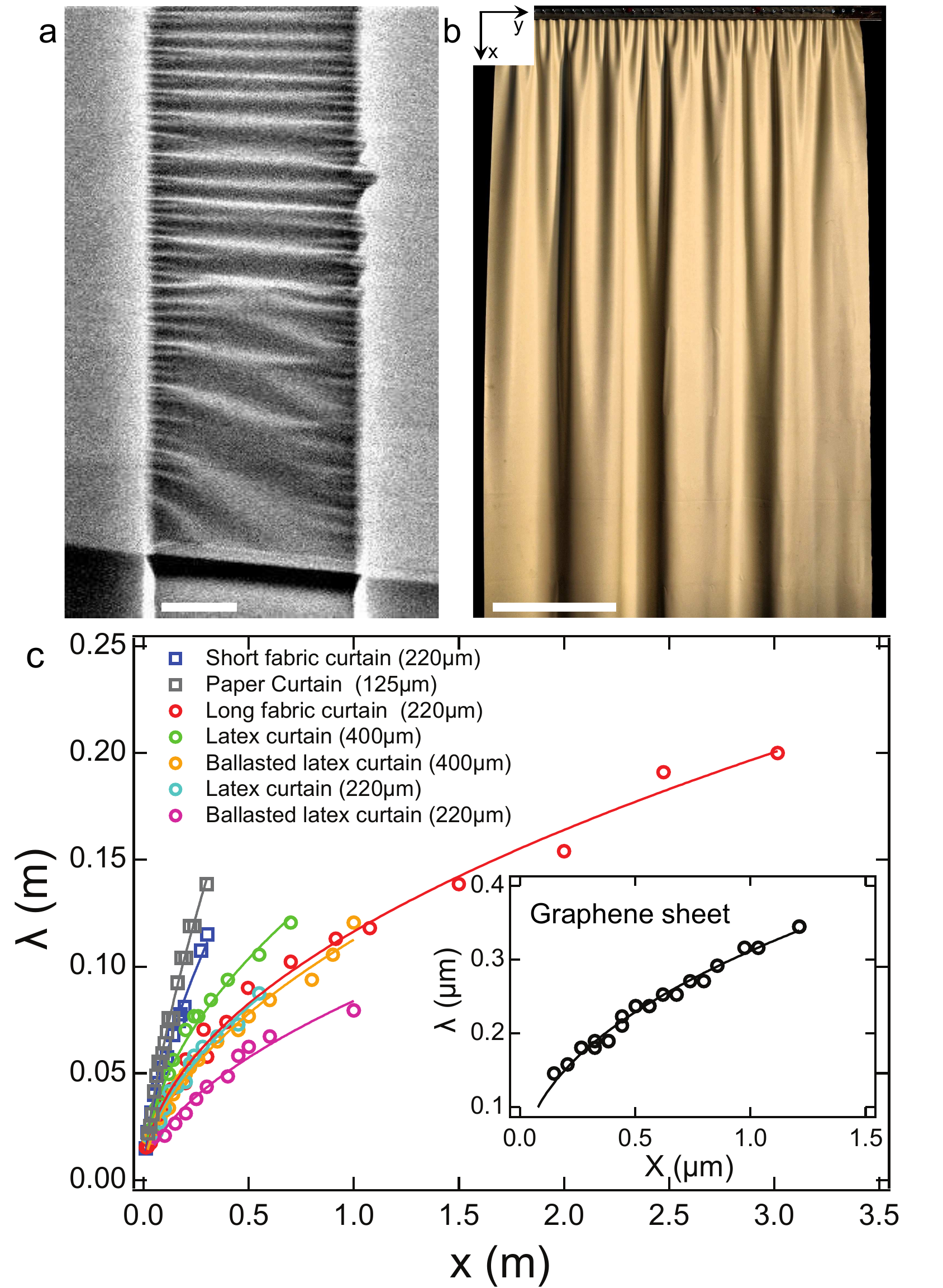}
\vspace{-2mm}
\caption{
{\bf a}, SEM image of a graphene bilayer thin sheet suspended across pre-defined trenches on Si/SiO$_2$ substrates (scale bar: $1\,\mu$m).
{\bf b}, Hierarchical pattern of folds obtained for a long suspended curtain made of a thin sheet of rubber (scale bar: 25 cm). 
{\bf c}, Evolution of the average wavelength, $\lambda$, with the distance from the constrained edge, $x$, for various curtains as indicated in the legend. Power law fits are added (the power exponents, $m$, are close to $2/3$ for the short fabric and the paper curtains and $1/2$ for the long fabric and the rubber curtains). Inset: Evolution of $\lambda$ with $x$ for the graphene sheet ($m$ is equal to 0.45 $\pm$ 0.02). The experimental parameters are detailed in the supplementary information. 
}
\label{fig1}
\end{figure}

To illustrate this hierarchical pattern, we show in Fig.~\ref{fig1}a wrinkled graphene sheet along with an ordinary hanged curtain.
These patterns are also similar to the self-similar circular patterns first reported by Argon {\it et al.} for the blistering of thin films adhering on a thick substrate~\cite{argon}.  
The diversity and complexity of those systems, characterized by various chemical and physical conditions, could suggest, {\it a priori}, that the underlying mechanisms governing the formation of these patterns are unrelated.
However, these systems can be depicted, independently from the details of the experiments, as a thin sheet constrained at one edge while the others are free to adapt their morphology. These constraints can take the form of an imposed wavelength at one edge or just the requirement that it should remain flat. 

\begin{figure}
\includegraphics[width=0.8\columnwidth]{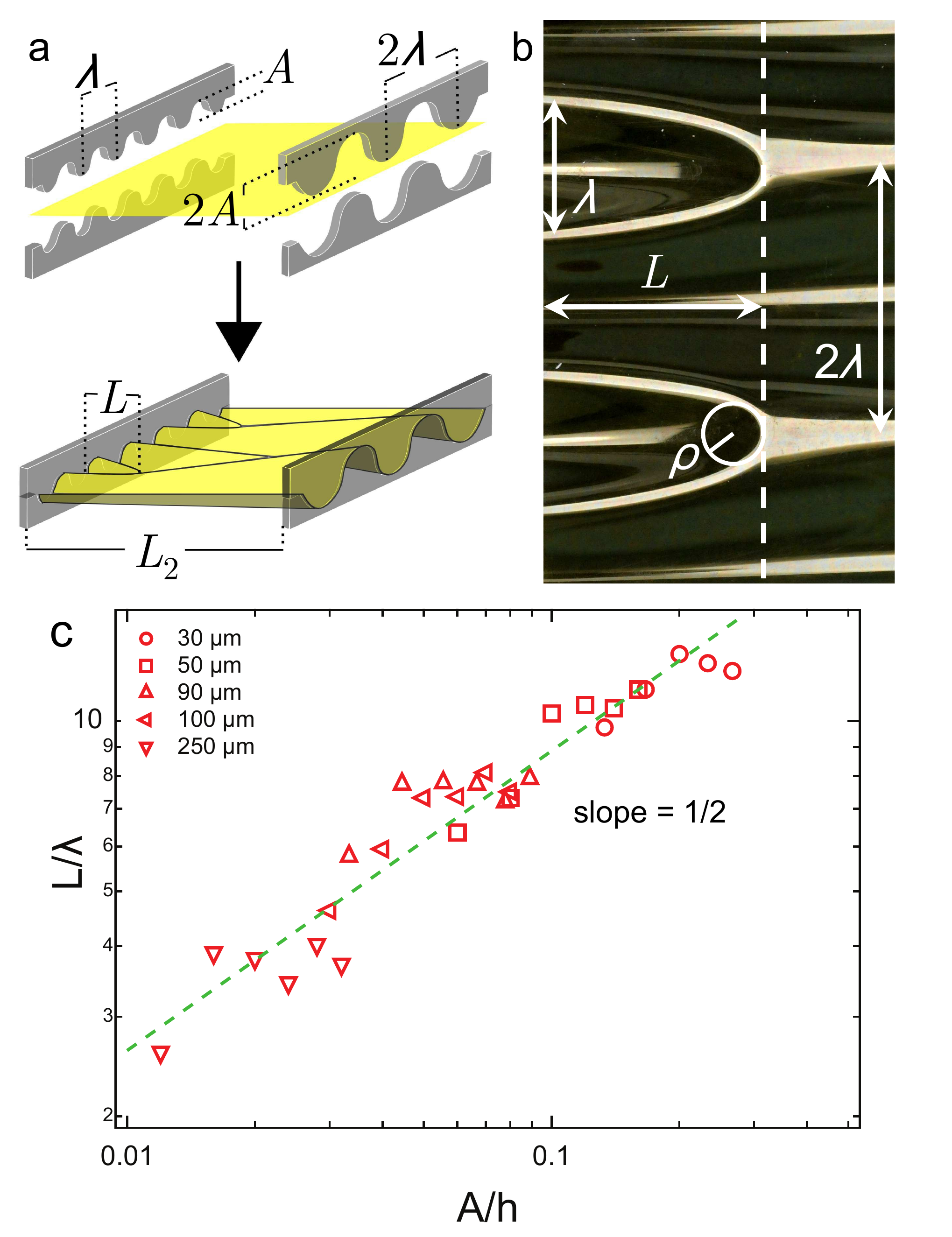}
\vspace{-2mm}
\caption{ {\bf a}, Schematic representation of the wrinklon experiments.
{ \bf b}, Morphology of the transition $\lambda$ to $2\lambda$ for a constrained plastic sheet for $A=6$ mm and $\lambda=8$ mm.
{ \bf c}, Evolution of the normalized length of a wrinklon, $L/\lambda$, with the normalized amplitude, $A/h$  (fixed wavelength, $\lambda=8$ mm) for different thicknesses as indicated. 
}
\label{fig2}
\end{figure}

As illustrated in Fig.~\ref{fig1} and \ref{fig2}, sheets made from various materials constrained at one edge by an imposed sinusoidal profile spontaneously develop a hierarchical pattern of folds or wrinkles.
At first sight, as quoted by numerous authors~\cite{argon,gioia,conti,sternberg,audo10,menon10,benny09,cerda04}, these patterns consist of a hierarchy of successive generations of folds whose typical size gradually increases along $x$ (Fig.~\ref{fig1}b). 
We propose to rationalize these various hierarchical patterns by considering the evolution of the average wavelength, $\lambda$, with the distance to the constrained edge, $x$.
This evolution is adequately described by a simple power law, $\lambda \sim x^{m}$, see Fig.~\ref{fig1}c, which confirms the self-similarity of these patterns as hypothesized in previous theoretical studies \cite{gioia,conti,sternberg,audo10}. Interestingly, curtains made of various materials with contrasted properties exhibit similar exponents. We observe values close to $2/3$ for ``light'' sheets and to $1/2$ for ``heavy'' sheets. Therefore the exponent $m$ is a robust feature of these folding patterns.

In this work, we describe in terms of simple scaling laws the theoretical arguments developed in the mathematical studies of the von Karman equation~\cite{conti,sternberg,audo10,benny09} and infer the properties of the hierarchical patterns. We also compare these results with extensive experimental data. To the best of our knowledge, the experimental characterization of these patterns had not been carried out before.

Assuming inextensibility of the sheet along the $y$ direction, the imposed undulation along this direction exactly compensates for an effective lateral compression of the membrane by a factor $(1-\Delta)$ defined as, 
$(1-\Delta) \equiv W/W_0 = W / \int_0^{W} \sqrt{1 + (\partial z/\partial y)^{2}} \, {\rm d}y$,
where $W_0$ and $W$ are the  curvilinear  and projected  width of the curtain, respectively, and $z(x,y)$ is the out-of-plane deformation of the sheet. At any position along the $x$ axis, the function $z(x,y)$ is typically sinusoidal along $y$, with an amplitude $A(x)$ and a wavelength $\lambda(x)$. 
The inextensibility hypothesis along the $y$ axis imposes
$\Delta \sim (A/\lambda)^2$ at the lowest order, where the lateral compression is assumed to be constant throughout the length of the curtain. The undulations of the sheet along $y$ are characterized by a curvature $\kappa \simeq \partial^2 z / \partial y^2$ whose typical value, varying along $x$, is of order $\kappa(x) \sim A/\lambda^2$. The corresponding energy per unit area, $u_{\rm b}$, for bending the membrane is thus of order $u_{\rm b}\sim Eh^3\,\kappa^2  \sim Eh^3 \Delta/\lambda^2$, 
where $E$ is the Young modulus and $h$ the thickness of the sheet. Since $u_{\rm b}$ is proportional to $1/\lambda^2$, the membrane adopts the largest possible wavelengths, in order to minimize energy. This tendency to increase the wavelength, combined with the constraint imposed at the boundaries, is the source of the observed hierarchical wrinkling pattern. 

Inspired by previous models based on successive period-doubling transitions~\cite{sternberg,audo10,benny11}, we consider that the allometric laws mentioned above can be derived by considering that the global pattern results from the self-assembly of building-blocks which we denote \emph{wrinklons}. 
A single wrinklon corresponds to the localized transition zone needed for merging two wrinkles of wavelength $\lambda$ into a larger one of width $2\lambda$. This transition requires a distortion of the membrane which relaxes over a distance $L$. In other words, each wrinklon is characterized by a size, $L$, which depends on the material properties and on the wavelength $\lambda$. 
To investigate the properties and behavior of wrinklons, we have performed model experiments using thin plastic sheets. The sheets were constrained with sinusoidal clamps: two opposite edges are constrained by a wavelength $\lambda$ (amplitude $A$) and $2\lambda$ (amplitude $2A$), respectively, see Fig.~\ref{fig2}a,b. 
The normalized size of the wrinklons, $L/\lambda$, is plotted in Fig.~\ref{fig2}c as a function of the normalized amplitude, $A/h$; the data collapse on a single curve defined by $L/\lambda \sim \sqrt{A/h}$. This relation implies that $L\propto \lambda^{3/2}$ since $A \sim \lambda \sqrt{\Delta}$. 

In a further step, the wrinklons can be assembled to mimic the behavior of a complete hierarchy. Indeed, if $L$ is the distance over which the wavelength increases from $\lambda$ to $2\lambda$, its variation, ${\rm d}\lambda/{\rm d}x$, is thus of order $\lambda/L$. Hence, the evolution of $\lambda$ as a function of the distance from the constrained edge, $x$, is given by 
\begin{equation}
\label{Eq:defLdlambdadx}
\frac{{\rm d}\lambda}{{\rm d}x} \simeq \frac{\lambda}{L}.
\end{equation}
Considering the scaling $L\propto \lambda^{3/2}$ deduced from the single wrinklon experiments, equation~(\ref{Eq:defLdlambdadx}) indicates that the wavelength along the sheet should evolve like $\lambda \propto x^{2/3}$. The excellent agreement between this power law and the experimental data measured for light sheets (Fig.~\ref{fig1}c) provides a strong support to the concept of wrinklons as building-blocks.
Equation~(\ref{Eq:defLdlambdadx}) can now be regarded as a tool that connects the properties of single wrinklons to the features of the full wrinkling-cascade pattern. 

We now focus on the  description of an elementary building block. For confined thin sheets, stretching deformations are costly as compared to pure bending. The sheet tends to adopt an isometric (developable) shape~\cite{Witten07}. However, the only developable solutions compatible with boundary conditions generally include flat domains surrounded by edge or point-like singularities. These singularities, which focus the elastic energy into narrow regions, have been classified as developable cones~\cite{Benamar,cerda99}, ridges~\cite{witten97,Witten07}, or curved ridges~\cite{pogorelov}. 
In our case, the scenario is however significantly different: in contrast to crumpling, stretching is smoothly distributed in the transition zone as pointed out recently in numerical simulations of deformed membranes~\cite{benny11}. 
The necessary stretching required for connecting the periodic patterns can be illustrated by a simple origami model made with a sheet of paper (see supp. info.).
The stretching energy can be estimated through the elongation strain of the sheet along $x$ within a transition domain.
The typical value of the strain along $x$ is of order $\alpha^2$, where $\alpha \sim A/L \sim \lambda \Delta^{1/2}/L$ is the average slope of the membrane.
The stretching energy thus reads 
$U_{\rm s} \sim Eh\,(\alpha^2)^2 L\lambda \sim Eh\,\lambda^5 \Delta^2 L^{-3}.$

As observed in Figs.~\ref{fig1} and \ref{fig2}, wrinklons should also include a tip singularity (a small region where Gaussian curvature is large). This singularity can be described as a semi-circular fold of radius $\rho$ (Fig.~\ref{fig2}b). The energy of these singularities has been derived by Pogorelov~\cite{pogorelov} in a study of deformed shells. In our context, the energy of such curved folds reads 
$U_{\rm cf} \sim Eh^{5/2}\,\alpha^{5/2}\rho^{1/2} \sim Eh^{5/2}\,\Delta^{5/4} \lambda^{7/2} L^{-3},$
where the radius at the tip of the wrinklon is taken as $\rho \sim \lambda^2/L$ as  suggested by the roughly parabolic shape of the crest of the defect (Fig.~\ref{fig2}b). 
Nevertheless, the ratio of the curved fold energy to the stretching energy of the wrinklon, $U_{\rm cf}/ U_{\rm s} \sim (h/A)^{3/2}$, is very small in our experiments : the effect of this concentrated region can therefore be neglected in the following.

\begin{figure}
\includegraphics[width=0.76\columnwidth]{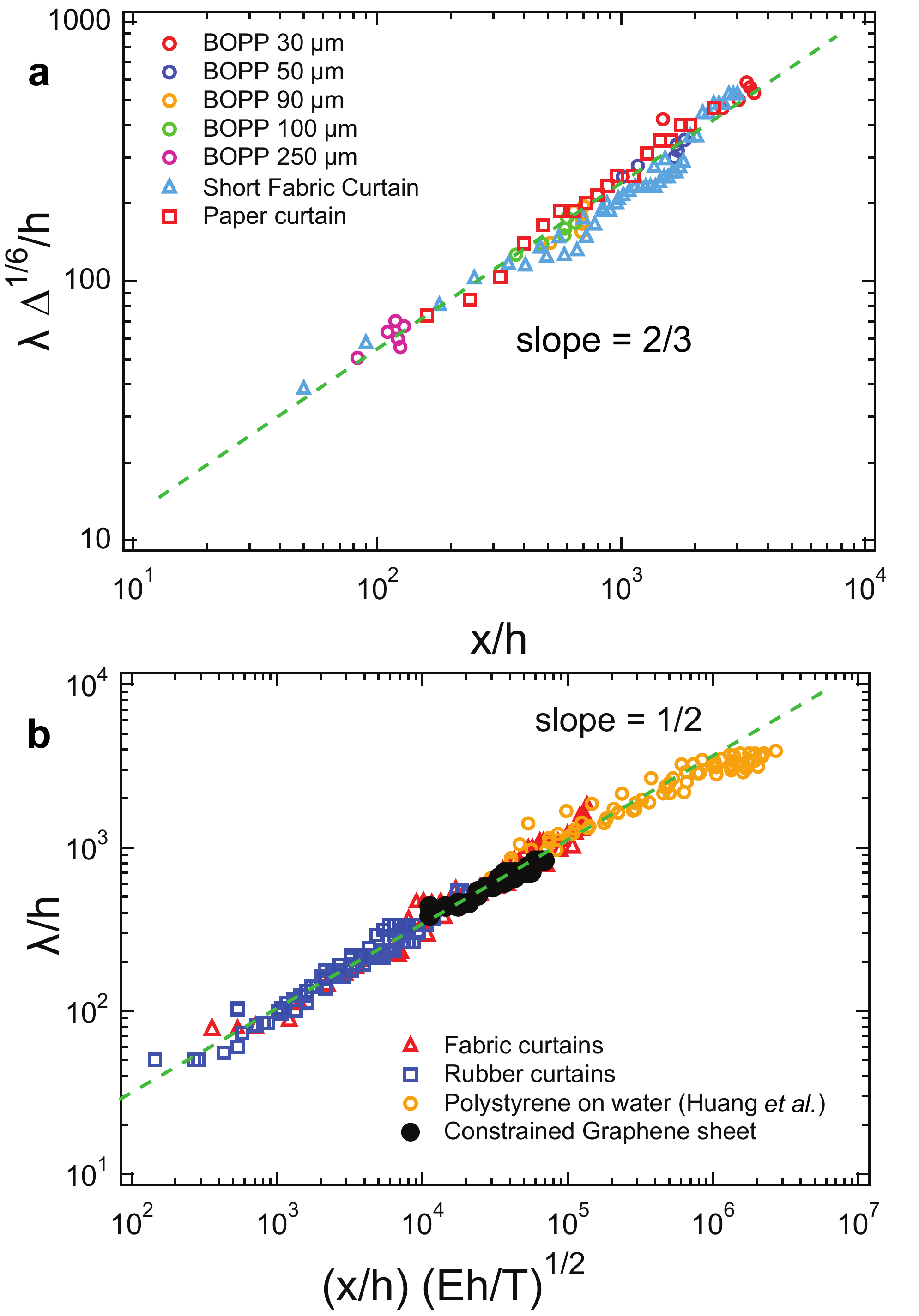}
\vspace{-2mm}
\caption{Master curves gathering all data. {\bf a}, Normalized wavelength, $\tilde{\lambda} = \lambda \Delta^{1/6}/h$, as a function of the normalized distance, $\tilde{x}= x/h$, from the constrained edge for short light sheets (fabric curtain, paper curtain and constrained plastic sheets). 
Dashed line: $\tilde{\lambda} = 2.89\ \tilde{x}^{0.65}$.
{\bf b}, Normalized wavelength $\tilde{\lambda} = \lambda/h$ as a function of the normalized distance from the constrained edge $\tilde{x}= (x/h)(Eh/T)^{1/2}$ for sheets under tension: fabric curtains, rubber curtains, suspended bilayer graphene sheet and polystyrene thin films deposited on water from Ref.~\cite{menon10}. 
Dashed line: $\tilde{\lambda} = 2.85\ \tilde{x}^{0.52}$.
}
\label{fig4}
\end{figure}

The total energy of a wrinklon, of characteristic area $L\lambda$, is thus given by $U_{\rm tot} = U_{\rm s} + U_{\rm b}  \simeq  Eh\,\lambda^5 \Delta^2 L^{-3} + Eh^3 \Delta L \lambda^{-1}$.
The size of a single wrinklon is finally obtained by minimizing $U_{\rm tot}$ with respect to $L$, yielding 
\begin{equation}
\label{Eq:Llambda32}
L(\lambda) \sim \Delta^{1/4}\ \lambda^{3/2}h^{-1/2}.
\end{equation}
This scaling emerges from a balance between bending and stretching energies and was previously reported for other situations, such as the decay length of an imposed curvature in a sheet~\cite{witten97} or the extension of a pinch in a pipe~\cite{maha07}. 
The scaling for the wavelength describing the whole hierarchical pattern is obtained by integration of equation~(\ref{Eq:defLdlambdadx}) with $L(\lambda)$ given by equation~(\ref{Eq:Llambda32}) and is found to be
\begin{equation}
\label{Eq:lambdax23}
\frac{\lambda(x) \Delta^{1/6}}{h} \sim \left(\frac{x}{h}\right)^{2/3}.
\end{equation}
The scaling law, $\lambda \propto x^{2/3}$, is in very good agreement with the observed power laws for light curtains, {\it e.g.} made of fabric or paper sheets (Fig.~\ref{fig1}c). 
In addition to yielding the proper exponent, this relation enables the comparison of the data obtained from seemingly disparate systems, over a wide range of lengthscales and independently of material properties. Figure~\ref{fig4}a provides a remarkable collapse of the evolutions of the wavelengths measured with light curtains and various thin plastic sheets.

Heavy curtains, made from fabric or rubber, and constrained graphene bilayers do not follow this behaviour (instead, they obey $\lambda \propto x^{1/2}$). 
In these experiments, an additional tensile force is acting on the sheet. This tension, $T$, is given by the longitudinal tensile strain induced by thermal manipulation in the case of graphene sheets~\cite{bao09} and by gravity  for heavy curtains ($T= \rho_c g h (H-x) \sim \rho_c g h H$, where $\rho_c$, $g$, $h$, and $H$ are the density of the curtain, the gravity constant, the thickness and the height of the curtain). These systems can also be compared to the cascade of wrinkles observed for compressed thin polystyrene films on an air-water interface~\cite{menon10} since the surface tension of water at the free edges pulls the thin sheet.

The tension exerted along $x$ per unit width, imposes an additional stretching energy given by
$U_{\rm t} \sim T\, \alpha^2\ L\lambda \sim T \Delta \lambda^3 L^{-1},$
and becomes dominant when $U_{\rm t} > U_{\rm s}$, that is when $T > Eh^2\Delta/A$. The total energy of the distorted membrane thus becomes ${ U}_{\rm tot} = { U}_{\rm t} + {U}_{\rm b}$. The length of a wrinklon found from the minimization of ${ U}_{\rm tot}$ is
\begin{equation}
\label{Eq:Llambda2}
L(\lambda)\sim \frac{\lambda^2}{h}\sqrt{\frac{T}{Eh}}.
\end{equation}
Similar relations, reflecting a balance between tension and bending energies, were previously proposed for single wavelength patterns in stretched sheets and heavy curtains~\cite{cerd03,cerda04}.
As expected, the tensile force increases the length of wrinklons for a given wavelength~\cite{benny09}. By integration of equation~(\ref{Eq:defLdlambdadx}) with $L(\lambda)$ given by equation~(\ref{Eq:Llambda2}), we obtain the corresponding spatial evolution of the wavelength along a heavy sheet
\begin{equation}
\label{Eq:lambdax12}
\frac{\lambda(x)}{h} \sim \left(\frac{Eh}{T}\right)^{1/4}\,\left(\frac{x}{h}\right)^{1/2}.
\end{equation}
This scaling is in excellent agreement with the power laws observed for heavy curtains and graphene bilayers, (Fig.~\ref{fig1}c). 
The data of various macroscopic curtains, graphene bilayers and nanometric polystyrene indeed collapse onto a single master curve without any fitting parameters (see Fig.~\ref{fig4}b). 
Our formalism is thus validated from hundreds of nm for graphene sheets to meters for rubber and fabric curtains, which shows the universality of our description.
The transition between the stretching and tension regimes can be obtained by comparing the relations (\ref{Eq:lambdax23}) and (\ref{Eq:lambdax12}). The critical distance from the edge at which this transition occurs is given by $x^{\star}/h \sim (Eh/T)^{3/2}\Delta$. In gravity dominated systems, the tension $T \simeq \rho g h H$,
gives the typical curtain length $H_c \sim h (E/\rho g h)^{3/5}\Delta^{2/5}$ above which  tension dominates.
Curtains shorter than $H_c$  (about 1m for our fabric) were used to observe the  regimes dominated by stretching (``light sheets"), whereas the top part of longer curtains were used for experiments concerning ``heavy sheets".

In summary, we showed that the self-similar patterns observed in sheets constrained at one edge cannot be described with d-cone or ridges singularities. In contrast, they can be built by stitching together building-blocks, called wrinklons characterized by a diffuse stretching energy. The self-similar structure is then related to the size of these wrinklons that depends on material properties and the local wavelength.  
Interestingly, we also show that these building-blocks can be readily manipulated through the size and energy cost of a single wrinklon by applying a tension. For large values of tension, we even expect a transition towards a purely cylindrical pattern along the sheet with a single wavelength. 
Finally, we can draw a parallel between this study and the previously reported fractal buckling of torn plastic sheets~\cite{sharon}. The imposed metric indeed determines the three-dimensional shape of the distorted membrane, characterized by a superimposition of various modes. In contrast, the patterns observed here for constrained thin sheets exhibit a continuous evolution of the wavelength.

{\bf Acknowledgements}  

The authors thank T. Witten , B. Davidovitch, N. Menon for fruitful discussions. This work was partially supported by the Belgian National Funds for Scientific Research (FNRS), 
the Government of the Region of Wallonia (REMANOS Research Programs), the European Science Foundation (Eurocores FANAS, EBIOADI), the French ANR MecaWet and the MIT-France MISTI program. C.N.L. and W.B. acknowledge the support by ONR N00014-09-1-0724 and the FENA Focus Center. The theoretical part of this work was mostly completed at the Aspen Center for Physics.

\clearpage

{\bf {\it Supplementary information for} ``Wrinkling hierarchy in constrained thin sheets from suspended graphene to curtains''}

\section{Methods}

\subsection{Curtains} 

The curtain experiments were carried out using fabric: thickness $220$ $\mu$m, density $\rho_c=820$ kg/m$^3$ and elastic modulus $E \sim 1$ MPa; natural latex:  thicknesses $220$ and $400$ $\mu$m, density $\rho_c=980$ kg/m$^3$ and elastic modulus $E \sim 1$ MPa; paper: thickness $125$ $\mu$m. The fabric curtain was 4 m long and 2.5 m wide, while the rubber and paper curtains were 2 m long and 1 m wide. Each curtain was hung with an imposed sinusoidal boundary condition through an array of screws of tunable length. 
The wavelength was fixed to $\lambda=20$ mm with an amplitude $A$ spanning  from 1.2 to 8 mm. The bottom edges of the curtains were left free except for the ballasted curtain experiments where 10 steel disks of $180$ g each were hung from the bottom edge with a spacing of $100$ mm in between. The average wavelength was obtained by dividing the total width of the curtain by the number of wrinkles found at a given distance from the constrained edge.   
For the various curtains, the evolution of $\lambda$ with the distance from constrained edge was analyzed with power laws. These fits yield the following exponents:
\begin{itemize}
\item [-] 0.62 $\pm$ 0.02, short fabric curtain; 
\item [-] 0.66 $\pm$ 0.07, paper curtain; 
\item [-] 0.50 $\pm$ 0.01, long fabric curtain; 
\item [-] 0.51 $\pm$ 0.02, $400\, \mu$m rubber curtain; 
\item [-] 0.53 $\pm$ 0.03, $220\, \mu$m rubber curtain; 
\item [-] 0.54 $\pm$ 0.04, ballasted $220\, \mu$m rubber curtain; 
\item [-] 0.53 $\pm$ 0.02, ballasted $400\, \mu$m rubber curtain. 
\end{itemize}

\subsection{Graphene bilayers} 

Suspended graphene membranes were prepared by standard mechanical cleavage technique on Si/SiO2 wafers with pre-patterned trenches. Bilayer graphene sheets were identified by color contrast in an optical microscope and/or Raman spectroscopy described in W. Bao {\it et al.} Nature Nanotech. {\bf 4}, 562 (2009).
The Si substrates were p-doped, and the thickness of silicon and SiO2  are 0.5 mm and 300 nm, respectively. The trenches, fabricated at the UCSB Nanofabrication facility, were defined by photolithography followed by plasma etching in a reactive ion etcher (RIE) system. Strain in graphene sheets was estimated using the same procedure as the one used to produce Fig.~1e in W. Bao {\it et al.} (2009).

SEM image of graphene bilayer thin sheets suspended across pre-defined trenches on Si/SiO$_2$ substrates and annealed at 523K were recorded with a tilt angle of $75^{\circ}$ to improve the contrast in amplitude.

\subsection{Isolated wrinklons.} 

These experiments were carried out using A4 size sheets made of biaxially oriented polypropylene (BOPP Innovia) of thicknesses 30, 50, 90, 100 and 250 $\mu$m and elastic modulus $E \sim 2.6$ GPa.  To impose a sinusoidal condition at the edge of the sheet we used Plexiglas clamps of thickness 5 mm with sinusoidal profiles of different amplitudes and wavelengths to clamp the sheets: fixed amplitude $A=3$ mm with wavelengths $\lambda=13,\,18,\,26$ and $58$ mm; fixed wavelength $\lambda=24$ mm with amplitudes $A=3,\,4,\,5,\,6$ and $7$ mm. A second clamp of wavelength $2\lambda$ and amplitude $2A$ was used to impose a second profile condition for the sheet at a certain distance $L_2$ away from the clamped edge. In this way we forced the sheet to switch from the initial wavelength $\lambda$ to the next wavelength $2\lambda$, thus generating the formation of a controlled wrinklon. 
The distance $L_2$ was selected by separating progressively the opposite clamps until the shape of the wrinkles became stationary and before a new generation of wrinklons could arise. 

\subsection{Morphology of wrinklons.}

\begin{figure*}
\includegraphics[width=\textwidth]{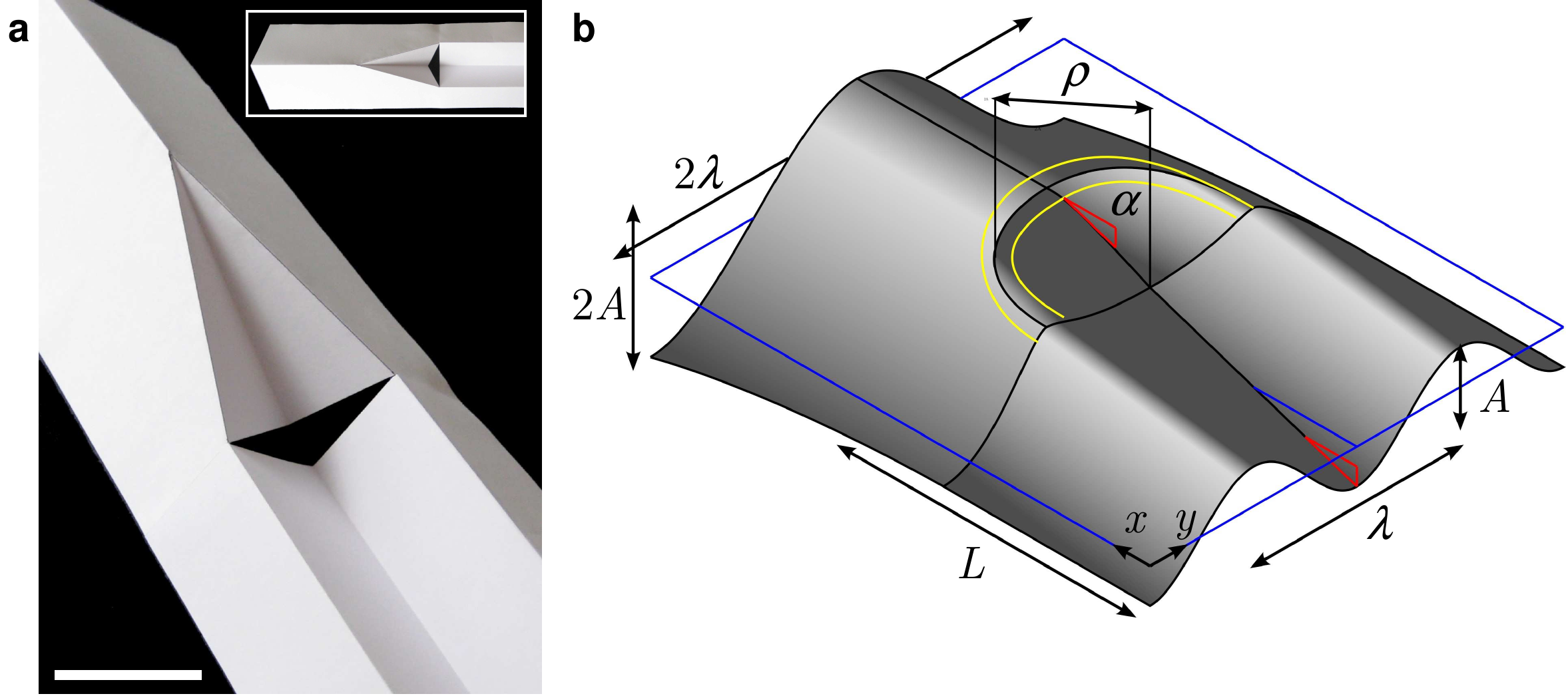}
\caption{{\bf a}, Origami describing the distorted domain at the $\lambda$ to $2\lambda$ transition. This toy model is illustrative of a sheet with a vanishing thickness and shows the necessity to stretch the sheet to connect both $\lambda$ and $2\lambda$ patterns. For sheets with vanishing thickness, the deformation energy is concentrated in sharp singularities (ridges or d-cones). The width/size of these singularities decreases as the thickness tends to zero. The limit would then correspond to sharp folds that is modeled by simple origami. The scale bar is 5 cm. {\bf b}, Schematic representation of the transition domain for sheets with finite thickness. The slope, $\alpha$, along $x$ direction is indicated.
}
\end{figure*}

\end{document}